\documentclass[a4paper,12pt, epsfig]{article}
\def\letter{0}\def\pr{0}
\pdfoutput=1 
\usepackage{epsfig}
\usepackage{epstopdf}
\usepackage{textcomp}
\usepackage{graphicx}
\usepackage{ifthen}
\usepackage{ulem}

\usepackage{amsmath}
\usepackage[psamsfonts]{amssymb}
\usepackage{euscript}

\usepackage{latexsym}
\usepackage[arrow,matrix,curve]{xy}

\usepackage[hypertexnames=false]{hyperref}
\hypersetup{
    colorlinks=true,
    linkcolor=blue,
    filecolor=magenta,   
    citecolor=black,   
    urlcolor=cyan,
    }

\newskip\humongous \humongous=0pt plus 1000pt minus 1000pt

\newif\ifdtup

\def\,{\hspace{-.1cm}}
\def\hsp{,\hspace{.7cm}}

\def\ddp#1{d^3\vp_{#1}}
\def\dvx{\int d^3\vx}

\def\ddpp#1{d^3\vpp_{#1}}

\def\fc#1#2 {\frac{n}{q}#1\frac{n}{q}#2}

\def\kt{\kappa}
\def\kt{\mathfrak{K}}
\def\ks{|\kt\rangle}

\def\hp{H^{\prime(t)}}
\def\rv{{\rm{vac}}}

\def\bp{\mathbf{P}}

\newcommand{\vac}{\ensuremath{|0\rangle}}
\newcommand{\ovac}{\ensuremath{|\Omega\rangle}}
\newcommand{\ps}{\ensuremath{|\vp_0\rangle}}

\renewcommand{\sin}{\textrm{sin}}

\renewcommand{\sinh}{\textrm{sinh}}
\renewcommand{\cosh}{\textrm{cosh}}
\renewcommand{\tanh}{\textrm{tanh}}
\newcommand{\sech}{\textrm{sech}}
\newcommand{\csch}{\textrm{csch}}

\def\exp#1{\hbox{\rm exp}\left[#1\right]}

\def\sign#1{{\rm sign}\left(#1\right)}

\renewcommand{\theequation}{\arabic{section}.\arabic{equation}}
\renewcommand{\(}{\begin{equation}}
\renewcommand{\)}{end{equation} \vspace{-.05in}\linebreak}

\newcounter{saveeqn}
\newcounter{savealpheqn}

\newcommand{\alpheqn}{\setcounter{saveeqn}{\value{equation}}%
  \stepcounter{saveeqn}\setcounter{equation}{0}%
  \renewcommand{\theequation}{\mbox{\arabic{section}.\arabic{saveeqn}
\alph{equation}}}
  \renewcommand{\)}{\end{equation}}}
\def\part#1{\frac{\partial}{\partial{#1}}}%
\def\group#1{\refstepcounter{equation}\setcounter{saveeqn}
 {\value{equation}}%
  \label{#1}\setcounter{equation}{0}%
\renewcommand{\theequation}{\mbox{\arabic{section}.\arabic{saveeqn}
\alph{equation}}}
  \renewcommand{\)}{\end{equation}}}
\newcommand{\reseteqn}{\setcounter{equation}{\value{saveeqn}}%
  \renewcommand{\theequation}{\arabic{section}.\arabic{equation}}%
  \renewcommand{\)}{\end{equation}}}

\newcommand{\aalpheqn}{\setcounter{saveeqn}{\value{equation}}%
  \stepcounter{saveeqn}\setcounter{equation}{0}%
  \renewcommand{\theequation}{\mbox{
        \Alph{subsection}.\arabic{saveeqn}\alph{equation}}}
   \renewcommand{\)}{\end{equation}}}
\newcommand{\areseteqn}{\setcounter{equation}{\value{saveeqn}}%
  \renewcommand{\theequation}{\Alph{subsection}.\arabic{equation}}%
  \renewcommand{\)}{\end{equation}}}

\renewcommand{\thefootnote}{\alph{footnote}}
\renewcommand{\(}{\begin{equation}}
\renewcommand{\)}{\end{equation}}
\newcommand{\ba}{\begin{eqnarray}}
\newcommand{\ea}{\end{eqnarray}}
\renewcommand{\a}{\alpha}
\renewcommand{\b}{\beta}

\renewcommand{\sl}{{\sqrt{\lambda}}}

\newcommand{\cbp}{\mathop{\vtop{\ialign{##\crcr
   $\hfil\displaystyle{}\hfil$\crcr\noalign{\kern-13pt\nointerlineskip}
   \BIG{)}\hskip0pt\crcr\noalign{\kern3pt}}}}}
\newcommand{\pa}{\mathop{\vtop{\ialign{##\crcr

$\hfil\displaystyle{\oplus}\hfil$\crcr\noalign{\kern+1pt\nointerlineskip
}
   \hspace{.08in}$^{\alpha=0}$\hskip6pt\crcr\noalign{\kern3pt}}}}}
\renewcommand{\hsp}{,\hspace{.3in}}
\newcommand{\p}{^\prime}
\newcommand{\pp}{^{\prime\prime}}

\catcode`\@=11
\def\vereq#1#2{\lower3pt\vbox{\baselineskip1.5pt \lineskip1.5pt
\ialign{$\m@th#1\hfill##\hfil$\crcr#2\crcr\sim\crcr}}}
\catcode`\@=12

\renewcommand{\(}{\begin{equation}}
\renewcommand{\)}{\end{equation}}

\def\vx{{\vec{x}}}
\def\vp{{\vec{p}}}
\def\vpp{{\vec{p}^{
\hspace{.05cm}\prime
}}}
\def\vppp{{\vec{p}^{
\hspace{.05cm}\prime\prime
}}}
\def\vk{{\vec{k}}}
\def\vkp{{\vec{k}\p}}

\def\pin#1{\int \frac{d#1}{2\pi}}
\def\ppin#1{\int\hspace{-17pt}\sum \frac{d#1}{2\pi}}
\def\ppink#1{\int\hspace{-17pt}\sum\frac{d^{#1}\vk}{(2\pi)^{#1}}}

\def\ppinkp#1{\int\hspace{-17pt}\sum\frac{d^{#1}k\p}{(2\pi)^{#1}}}
\def\dint{\int\hspace{-12pt}\sum }
\def\pink#1{\int \frac{d^{#1}k}{(2\pi)^{#1}}}

\def\pinkp#1{\int \frac{d^{#1}k\p}{(2\pi)^{#1}}}

\def\pinvp#1{\int \frac{d^{#1}\vp}{(2\pi)^{#1}}}
\def\pinvk#1{\int\hspace{-17pt}\sum \frac{d^{#1}\vk}{(2\pi)^{#1}}}
\def\kinv#1#2{\int\hspace{-17pt}\sum \frac{d^{#1}\vec{#2}}{(2\pi)^{#1}}}
\def\pinv#1#2{\int \frac{d^{#1}\vec{#2}}{(2\pi)^{#1}}}
\def\pinvx#1#2{\int d^{#1}\vec{#2}}
\def\pinvi#1#2#3{\int \frac{d^{#1}\vec{#2}_{#3}}{(2\pi)^{#1}}}
\def\ppinv#1#2{\int \frac{d^{#1}\vec{#2}^{
\hspace{.07cm}\prime
}}{(2\pi)^{#1}}}

\def\Bd#1{B^\ddag_{k_{#1}}}
\def\Ad#1{A^\ddag_{\vp_{#1}}}
\def\Btd#1{B^{(t)\ddag}_{k_{#1}}}
\def\Bt#1{B^{(t)}_{k_{#1}}}
\def\Bdp#1{B^\ddag_{k\p_{#1}}}

\def\cc{\mathcal{C}}
\def\df{\mathcal{D}_{f}}
\def\dv{\mathcal{D}_{v}}
\def\dft{\mathcal{D}_{f}^{(t)}}
\def\dfd{\mathcal{D}_{f}^{(t)\dag}}

\def\avkd{A^\ddag_{\vec{k}}}
\def\avpd{A^\ddag_{\vec{p}}}
\def\avkm{A_{-\vec{k}}}
\def\avpm{A_{-\vec{p}}}
\def\avk{A_{\vec{k}}}
\def\avp{A_{\vec{p}}}
\def\omvk{\omega_{\vec{k}}}
\def\omvp{\omega_{\vec{p}}}

\def\navkd#1{A^\ddag_{\vec{k}_{#1}}}
\def\navpd#1{A^\ddag_{\vec{p}_{#1}}}

\def\navpm#1{A_{-\vec{p}_{#1}}}
\def\navk#1{A_{\vec{k}_{#1}}}
\def\navp#1{A_{\vec{p}_{#1}}}

\def\nomvp#1{\omega_{\vec{p}_{#1}}}

\def\O#1#2{|\Omega\rangle_{#1}^{#2}}

\def\I{\mathcal{I}}

\def\os{\omega_S}

\def\gx{(\gamma(x-vt))}

\def\jhep#1{\textcolor{green}{#1}}
\def\red#1{\textcolor{red}{Jarah: #1}}
\def\blu#1{\textcolor{blue}{Hui: #1}}
\usepackage[dvipsnames]{xcolor}
\def\gre#1{\textcolor{ForestGreen}{Hengyuan: #1}}
\def\hui#1{\textcolor{Mulberry}{Hui: #1}}

\newcommand{\beas}{\begin{eqnarray*}}
\newcommand{\eeas}{\end{eqnarray*}}

\newcommand{\bquo}{\begin{quote}}
\newcommand{\enqu}{\end{quote}}

\def\lim#1{\stackrel{\rm{lim}}{{}_{#1}}}

\newcommand{\R}{{\mathbb R}}

    \newcommand{\g}{\mathfrak g}

\def\ch{{\mathcal{H}}}
\def\co{{\mathcal{O}}}

\def\ok#1{\omega_{k_{#1}}}
\def\okp#1{\omega_{k\p_{#1}}}
\def\ovp#1{\omega_{\vp_{#1}}}
\def\ovpp#1{\omega_{\vpp_{#1}}}
\def\ovppp#1{\omega_{\vppp_{#1}}}
\def\okt#1{\omega_{\kt_{#1}}}
\def\V#1{V^{(#1)}(\sqrt{\lambda}f(x))}
\def\Vg#1{V^{(#1)}(\sqrt{\lambda}f(\gamma(x-vt))}
\def\ck{\csch\left(\frac{\pi k}{2\b}\right)}

\def\mb{\mathcal{B}}
\def\mc{\mathcal{C}}
\def\md{\mathcal{D}}
\def\me{\mathcal{E}}
\def\gt{\tilde{\g}}
\def\dim{2}
\def\dimtwo{4}
\def\dimthree{6}
\def\phip{\phi}
\def\pip{\pi}
\def\bp{{\overline{p}}}

\def\Xint#1{\mathchoice
   {\XXint\displaystyle\textstyle{#1}}%
   {\XXint\textstyle\scriptstyle{#1}}%
   {\XXint\scriptstyle\scriptscriptstyle{#1}}%
   {\XXint\scriptscriptstyle\scriptscriptstyle{#1}}%
   \!\int}
\def\XXint#1#2#3{{\setbox0=\hbox{$#1{#2#3}{\int}$}
     \vcenter{\hbox{$#2#3$}}\kern-.5\wd0}}

\def\dashint{\Xint-}

\newcommand{\beq}{\begin{equation}}
\newcommand{\eeq}{\end{equation}}
\newcommand{\bea}{\begin{eqnarray}}
\newcommand{\eea}{\end{eqnarray}}

\newskip\humongous \humongous=0pt plus 1000pt minus 1000pt

\newif\ifdtup

\catcode`\@=11

\ifthenelse{\equal{\letter}{0}}{ 

\@addtoreset{equation}{section}
\def\theequation{\arabic{section}.\arabic{equation}}

\def\@normalsize{\@setsize\normalsize{15pt}\xiipt\@xiipt
\abovedisplayskip 14pt plus3pt minus3pt%
\belowdisplayskip \abovedisplayskip
\abovedisplayshortskip \z@ plus3pt%
\belowdisplayshortskip 7pt plus3.5pt minus0pt}

\def\small{\@setsize\small{13.6pt}\xipt\@xipt
\abovedisplayskip 13pt plus3pt minus3pt%
\belowdisplayskip \abovedisplayskip
\abovedisplayshortskip \z@ plus3pt%
\belowdisplayshortskip 7pt plus3.5pt minus0pt
\def\@listi{\parsep 4.5pt plus 2pt minus 1pt
      \itemsep \parsep
      \topsep 9pt plus 3pt minus 3pt}}

\relax

\def\section{\@startsection{section}{1}{\z@}{3.5ex plus 1ex minus  .2ex}{2.3ex plus .2ex}{\large\bf}}

\def\thesection{\arabic{section}}
\def\thesubsection{\arabic{section}.\arabic{subsection}}

\def\appendix{\setcounter{section}{0}
 \def\thesection{Appendix \Alph{section}}
 \def\thesubsection{\Alph{section}.\arabic{subsection}}
 \def\theequation{\Alph{section}.\arabic{equation}}}
\renewcommand{\theequation}{\arabic{section}.\arabic{equation}}

}{
\renewcommand{\theequation}{\arabic{equation}}

} 

\begin{document}
\def\thefootnote{\fnsymbol{footnote}}
\def\thetitle{An Extended Soliton's Zero Modes}
\def\autone{Jarah Evslin}
\def\authui{Hui Liu}
\def\autfour{Baiyang Zhang}
\def\auttwo{Hengyuan Guo}
\def\affyer{Yerevan Physics Institute, 2 Alikhanyan Brothers St., Yerevan, 0036, Armenia}
\def\affb{University of the Chinese Academy of Sciences, YuQuanLu 19A, Beijing 100049, China}
\def\affa{Institute of Modern Physics, NanChangLu 509, Lanzhou 730000, China}
\def\affc{School of Physics and Astronomy, Sun Yat-sen University, Zhuhai 519082, China}
\def\affe{Institute of Contemporary Mathematics, School of Mathematics and Statistics, Henan University, Kaifeng, Henan 475004, P. R. China}


\ifthenelse{\equal{\pr}{1}}{
\title{\thetitle}
\author{\autone}
\author{\auttwo}
\author{\autthree}
\affiliation {\affa}
\affiliation {\affb}
\affiliation {\affc}

}{

\begin{center}
{\large {\bf \thetitle}}

\bigskip

\bigskip


{\large 
\noindent  \autone{${}^{12}$}
\footnote{jarah@impcas.ac.cn} and 
\authui{${}^{3}$} \footnote{hui.liu@yerphi.am} 
}


\vskip.7cm

1) \affa\\
2) \affb\\
3) \affyer\\

\end{center}

}

\begin{abstract}
\noindent
Quantum field theory in the presence of localized solitons is more complicated than vacuum sector quantum field theory, largely as a result of the soliton's zero modes.  In the present work, we try to understand to what extent this situation carries over to extended solitons.  To this end, we explicitly construct the quantum state corresponding to a domain wall string or membrane, treating the zero modes carefully.  This is done both for compact and also infinitely extended solitons.  As an application, we calculate the differential probability for a domain wall string's translation mode, with a given momentum, to be excited by a single quantum of incoming radiation.  

\end{abstract}


%
\setcounter{footnote}{0}
\renewcommand{\thefootnote}{\arabic{footnote}}

\ifthenelse{\equal{\pr}{1}}
{
\maketitle
}{}

\section{Introduction}

The spectrum of small excitations about a soliton is different from that about an isolated vacuum, because the former enjoys a zero mode.  This zero mode invariably leads to additional interactions and additional states, and so, while many formalisms successfully treat quantum solitons \cite{sol1,sol2,sol3,sol4,cq1,cq2,gw22}, they each need to resolve the ``zero mode problem''  \cite{fkk75,fk76,mat77}. As a result, these formalisms are considerably more complicated than their vacuum sector counterparts.   The purpose of this note is to understand the role of zero modes in extended solitons.

Localized solitons in classical field theory can be lifted to quantum field theory, where they are quantum states \cite{skyrme}. The corresponding Hamiltonian eigenstates are translation-invariant, because their translation zero mode must be evenly spread throughout space.  This makes their wave functions nonnormalizable.  To calculate amplitudes, one must divide by the translation symmetry.  Such a quotient, needless to say, leads to a Jacobian factor that must be included in all amplitudes.

In the case of collective coordinate methods \cite{gs74,cl75,tom75}, this Jacobian factor is trivial.  While these methods continue to yield interesting results \cite{ito23,andy24}, nonetheless they have proven to be rather unwieldy as a result of the transformation to canonical coordinates and various limits \cite{gj76,bh24,ga24}.  Faster progress has been achieved recently with the more lightweight linearized soliton perturbation theory (LSPT) \cite{mekink,me2loop}.  However, in this case, the Jacobian factor is nontrivial~\cite{menorm}, leading to a higher order contribution to scattering.  While the Jacobian was derived independently of such considerations, this higher order contribution in fact turned out to be necessary in examples to get orthogonal Hamiltonian eigenstates \cite{menorm} and even for the translation-invariance of loop amplitudes in \cite{elastic}. 

Our main result, as summarized in Table \ref{tab}, will be a dramatic simplification in the case of extended solitons, with respect to localized solitons.  In the case of infinite, extended solitons the zero mode is just one member of a continuous family of translation modes.  In this case the zero mode wave function of a Hamiltonian eigenstate can be chosen to be normalizable.  This is not to say that translation invariance is always broken, in fact in the case of domain wall strings it is preserved by Coleman's theorem \cite{colemanth} applied to the (1+1)-dimensional string worldsheet theory, where it is an internal symmetry.  However, it is preserved as a result of the infrared yet nonzero translation modes.  This is quite different from the localized soliton, whose spectrum was gapped and whose translation-invariance resulted from the nonnormalizable zero mode.  As a result, in the case of infinitely extended solitons, no such quotient is necessary and the resulting corrections do not arise.

\begin{table}
\begin{center}
\begin{tabular}{|l|l|l|}
\hline
Soliton&Translation Invariance&Zero Mode\\
\hline\hline
kink&Preserved due to zero mode&Not Normalizable\\
\hline
finite string&Preserved due to zero mode&Not Normalizable\\
\hline
infinite string&Restored by nonzero modes&Normalizable\\
\hline
infinite membrane&Broken&Normalizable\\
\hline
\end{tabular}
\end{center}
\caption{Symmetry breaking patterns and zero mode wave function normalizability are reported for various solitons.  The zero mode is said to be normalizable if there is a ground state with a normalizable zero mode wave function.} \label{tab}
\end{table}

There are additional motivations for studying the zero modes in extended solitons.  For example, recently Ref.~\cite{hengyuan} has calculated the amplitude for Stokes scattering on the domain wall string.  This is the process in which an incoming perturbative meson scatters off of a domain wall string, exciting its shape mode.  The same calculation in principle could be applied to the excitation of the translation modes, describing how a domain wall string recoils when it is struck by radiation.  In fact, the relevance of such processes to cosmology has recently been highlighted in Refs.~\cite{cos1,cos2,cos3}.  However, without understanding how to incorporate the zero mode into the domain wall string's ground state, such an extension was not possible in Ref.~\cite{hengyuan}.  In the present paper we will describe the role of the zero mode in the ground state and use this to find the differential probability that incident radiation excites the translation modes as a function of the outgoing momenta.

We are also motivated by applications to nuclear physics.  At large $N$, nuclei are described by solitons \cite{skyrmion,wittskyrme,bjskyrme,rhoskyrme}.  On the other hand, the connection between experimental observations of nucleons and perturbative QCD is through factorization theorems \cite{fact} that separate amplitudes into perturbative and nonperturbative parts.  These factorization theorems only apply to the light front formulation, where the various distribution and fragmentation functions are defined.  However even the most basic questions regarding zero modes, such as whether they encode the nontrivial vacuum structure, are hotly debated on the light front \cite{zero0,zero1,zero2,zero3,zero4}.  This motivates us to try to understand soliton states, and in particular their zero modes, on the light front \cite{kinklf,kinklf2}.  The present paper is entirely in the instantaneous frame, but is meant to serve as a launch point for such an investigation.

Finally, in loop calculations involving solitons such as Ref.~\cite{elastic}, a prominent role is played by diagrams in which zero modes are created and destroyed.  These are rather unusual contributions, and make calculations much more complicated.  As the translations have infinite kinetic terms in the case of extended solitons, one may ask whether they decouple and so do not contribute to loop effects involving extended solitons.

We begin in Sec.~\ref{cysez} by quantizing a domain wall string \cite{zar98,dws1,dws2} wrapped around a circle of radius $R$.  Here, in the ground state, the zero mode is in a flat superposition implying that translation symmetry is unbroken.  Next, in Sec.~\ref{gsez} we consider the case in which $R$ is greater than any other scale in the problem.  Now we find there are degenerate ground states with arbitrary zero mode configurations.  In Ref.~\ref{infsez} we define the infinite domain wall string to be that of the previous section.  Note that although the zero mode can now have a nonzero wave function, translation symmetry is restored by the nonzero translation modes, in line with the usual intuition from Coleman's theorem.  This is generalized to higher dimensional domain walls in Sec.~\ref{memsez}, where the translation symmetry now may be broken.  While the symmetry breaking pattern is different, the zero mode wave function in higher dimensions is essentially the same as that of the string, which we have stressed is quite different from that of a localized soliton.  In Sec.~\ref{stsez} these developments are applied to calculate the differential probability of Stokes scattering, first for a general potential and then for the $\phi^4$ double-well model.

\section{The Domain Wall String on a Cylinder} \label{cysez}

We will begin with a (2+1)-dimensional scalar theory with one noncompact dimension $x\in (-\infty,\infty)$ and one compact dimension $y\in [-\pi R,\pi R)$.

\subsection{The Domain Wall String}

Consider a classical theory of a scalar field $\phi(x,y,t)$ and its dual momentum $\pi(x,y,t)$ described by the classical Hamiltonian density
\beq
\ch=\frac{\pi^2+\left(\partial_x\phi\right)^2+\left(\partial_y\phi\right)^2}{2}
+\frac{V(\sl\phi)}{\lambda}.
\eeq
Here the coupling, $\lambda$, is a parameter.  We will be interested in a potential $V$ with degenerate minima $\phi_1$ and $\phi_2$.  Furthermore, we assume that the second derivatives of the potential at the two minima are equal and we use them to define a positive mass $m$
\beq
m^2=V^{(2)}(\sl\phi_1).
\eeq

Consider a classical solution $\phi(x,y,t)$ corresponding to a domain wall string wrapping around the $y$ direction
\beq
\lim{x\rightarrow -\infty} {\phi(x,y,t)}=\phi_1\hsp
\lim{x\rightarrow \infty} {\phi(x,y,t)}=\phi_2\hsp 
\phi(x,y,t)=F(x,y)=f(x).
\eeq
Let the soliton be BPS, so that the string's classical energy is
\beq
Q_0=\int_{-\pi R}^{\pi R} dy \rho_0=2\pi R\rho_0\hsp
\rho_0=\int_{-\infty}^{\infty} dx f^{\prime 2}(x)
\eeq
where $\rho_0$ is the string tension.

A small, constant frequency perturbation about the string may be written
\beq
\phi(x,y,t)=f(x)+\g(x,y)e^{-i\Omega t}.
\eeq
A basis of solutions of the linearized equations of motion is given by perturbations that factorize
\beq
\g(x,y)=\g(x)e^{-ipy/R}
\eeq
where $p$ is integer so that $\g(x,y)$ will be single valued.  Here $\g(x)$ is a solution of the Sturm-Liouville equation
\beq
V^{(2)}(\sqrt{\lambda}f(x))\g(x)=\left(\omega^2+\partial_x^2\right)\g(x)\hsp \omega=\sqrt{\Omega^2-\frac{p^2}{R^2}}.
\eeq
This has three kinds of solutions, depending on $\omega$.  First, there will be a single zero mode
\beq
\g_B(x)=-\frac{f\p(x)}{\sqrt{\rho_0}}\hsp \omega_B=0
\eeq
where the sign is chosen for later convenience and the normalization is chosen so that $\int dx\  g_B^2(x)=1$.  Next, for every real $k$ there will be a continuum mode
\beq
\g_k(x)\hsp \omega_k=\sqrt{m^2+k^2}.
\eeq
Finally, there may be some discrete, real shape modes $\g_S(x)$ in the gap $0<\omega_S<m$.  We choose the normalization conditions
\beq
\int_{-\infty}^{\infty} dx \g_{k}(x)\g_{k\p}(x)=2\pi\delta(k+k\p)\hsp \int_{-\infty}^{\infty} dx \g_S^2(x)=1.
\eeq
The solutions of the Sturm-Liouville equation span the space of functions of $x$, while the plane waves $e^{-ipy/R}$ span the functions of $y$, and so our perturbations span the space of functions of $(x,y)$ and can be used to decompose Schrodinger picture quantum fields\footnote{More precisely, it is sufficient that a completeness relation is satisfied stating that there is a dual basis such that one can produce Dirac delta functions $\delta(x,y)$ by folding the normal modes $\g(x,y)$ into the dual modes.  The dual modes satisfying the completeness relation are just $\g^*(x,y)$.}.

Notice that $\Omega=0$ only in the case of $\g_B(x)$, corresponding to $k=B$ and $p=0$.  More generally we write
\beq
\Omega_{Bp}=\frac{|p|}{R}\hsp \Omega_{Sp}=\sqrt{\omega_S^2+\frac{p^2}{R^2}}\hsp \Omega_{kp}=\sqrt{m^2+k^2+\frac{p^2}{R^2}}.
\eeq

\subsection{The Operators}

Now let us pass to Schrodinger picture quantum field theory.  We will take $\lambda\hbar/m$ to be arbitrarily small and positive, defining a semiclassical expansion.  Then we set $\hbar=1$.  Also, the Hamiltonian will be
\beq
H=\int_{-\infty}^{\infty} dx\int_{-\pi R}^{\pi R} dy :\ch(x,y):
\eeq
where $::$ is the usual normal ordering.

We will consider a single, Schrodinger picture field quantum field\footnote{We use the same notation for the classical and the Schrodinger picture quantum fields, but the classical fields contain an additional argument $t$.} $\phi(x,y)$ and its dual momentum $\pi(x,y)$ which satisfy the canonical commutation relations
\beq
[\phi(x_1,y_1),\pi(x_2,y_2)]=i\delta(x_1-x_2)\delta(y_1-y_2). \label{ccr}
\eeq

Following Refs.~\cite{sol3,mekink,noi3d} we decompose the Schrodinger picture fields in terms of the domain wall string's normal modes
\beq
\phi(x,y)=\frac{1}{2\pi R}\ppin{k}\sum_{p=-\infty}^\infty \phi_{kp}\g_k(x)e^{-ipy/R}\hsp
\pi(x,y)=\frac{1}{2\pi R}\ppin{k}\sum_{p=-\infty}^\infty \pi_{kp}\g_k(x)e^{-ipy/R}
\eeq
where $\dint\frac{dk}{2\pi}$ is the sum of an integral $\pin{k}$ over real values of $k$ plus all discrete shape modes $S$ and also the zero mode $B$.  As these normal modes are a basis, the decomposition is invertible
\beq
\phi_{kp}=\int_{-\infty}^{\infty} dx\int_{-\pi R}^{\pi R} dy \phi(x,y)\g_{-k}(x)e^{ipy/R}\hsp
\pi_{kp}=\int_{-\infty}^{\infty} dx\int_{-\pi R}^{\pi R} dy \pi(x,y)\g_{-k}(x)e^{ipy/R}.
\eeq
We remind the reader that in our compact notation $p$ runs over all integers while $k$ can be real for a continuum mode, but also can be a discrete index $B$ or $S$ representing the zero mode or a shape mode.  In the case of discrete indices, since the normal modes are taken to be real, we define $\g_{-S}=\g_{S}$ and $\g_{-B}=\g_{B}$.

The canonical commutation relations (\ref{ccr}) then yield the commutation relations
\beq
[\phi_{k_1p_1},\pi_{k_2p_2}]=\left\{\begin{tabular}{lll}
${2\pi i R}{}\delta_{p_1,-p_2}\delta_{k_1k_2}$&&\rm{if $k_i$ are discrete}\\
${2\pi i R}{}\delta_{p_1,-p_2}\left(2\pi\delta(k_1+k_2)\right)$&&\rm{if $k_i$ are continuous}.
\end{tabular}
\right.
\eeq

Except for $\phi_{B0}$ and $\pi_{B0}$, we may reexpress these in terms of creation and annihilation operators
\beq
[B_{k_1p_1},B^\ddag_{k_2p_2}]=\left\{\begin{tabular}{lll}
$\delta_{p_1 p_2}\delta_{k_1k_2}$&&\rm{if $k_i$ are discrete}\\
$\delta_{p_1 p_2}\left(2\pi\delta(k_1-k_2)\right)$&&\rm{if $k_i$ are continuous}
\end{tabular}
\right.
\eeq
as
\beq\label{bddag1}
\phi_{kp}=\sqrt{{2\pi R}{}}\left(B^\ddag_{kp}+\frac{B_{-k,-p}}{2\Omega_{kp}}\right)\hsp
\pi_{kp}=i\sqrt{{2\pi R}{}}\left(\Omega_{kp}B^\ddag_{kp}-\frac{B_{-k,-p}}{2}\right).
\eeq
Here we have introduced the rescaled adjoint
\beq
B^\ddag_{kp}=\frac{B^\dag_{kp}}{2\Omega_{kp}}.
\eeq

\subsection{The Ground State}

We will now construct the domain wall string ground state $|K\rangle$ defined to be the minimum energy $Q$ eigenstate of the Hamiltonian
\beq
H|K\rangle=Q|K\rangle
\eeq
in the domain wall string sector.

First, we will decompose $|K\rangle$ into a perturbative piece $\vac$ acted upon by a nonperturbative displacement operator $\df$
\beq
|K\rangle=\df\vac\hsp
\df=\exp{-i\int_{-\infty}^{\infty} dx f(x) \int_{-\pi R}^{\pi R} dy \pi(x,y)}.
\eeq
The displacement operator $\df$ shifts $\phi(x,y)$ by the classical solution
\beq
\df^\dag \phi(x,y)\df=\phi(x,y)+f(x)
\eeq
and so takes the vacuum sector state $\vac$ into the domain wall string sector.

It is easy to show that $\vac$ is an eigenstate of the soliton Hamiltonian $H\p$
\beq
H\p\vac=Q\vac\hsp
H\p=\df^\dag H\df. \label{schrod}
\eeq

We will solve the eigenvalue equation (\ref{schrod}) order by order in $\lambda$.  We begin by decomposing $H\p$,\ $Q$ and their densities in powers of the coupling
\bea
Q&=&\sum_i Q_i\hsp \rho=\sum_i\rho_i\hsp
H\p=\sum_i H\p_i\hsp 
\ch\p(x,y)=\sum_i \ch\p_i(x,y)\nonumber\\
Q_i&\sim& O(\lambda^{i-1})\hsp \rho_i\sim O(\lambda^{i-1})\hsp H\p_i,\ \ch_i(x,y)\sim O(\lambda^{i/2-1}).
\eea
While the overall normalization of $\vac$ is arbitrary, we will also decompose $\vac$
\beq
\vac=\sum_i \vac_i\hsp \vac_i\sim O(\lambda^{i/2}).
\eeq

Using the definition (\ref{schrod}) one may evaluate
\beq
H\p_0=Q_0\hsp H\p_1=0\hsp
H\p_{n>2}=\lambda^{\frac{n}{2}-1}\int_{-\infty}^{\infty} dx \frac{V^{(n)}(\sqrt{\lambda} f(x))}{n !}\int_{-\pi R}^{\pi R} dy: \phi^n(x,y):.\label{hn}
\eeq
Note that the interaction terms $H\p_{n>2}$ are suppressed by powers of $\sl$ and so, to diagonalize $H\p$, one must first exactly, simultaneously diagonalize $H\p_0$, $H\p_1$ and $H\p_2$ and then treat $H\p_{n>2}$ using old fashioned perturbation theory.  $H\p_0$ is already diagonal as it is a $c$-number and $H\p_1$ vanishes.  Thus to begin the perturbation theory one must diagonalize $H\p_2$, which is
\bea
H\p_2&=&\int_{-\infty}^{\infty} dx\int_{-\pi R}^{\pi R} dy :\ch\p_2(x,y):
\\
\ch\p_2(x,y)&=&\frac{\pi^2(x,y)+\left(\partial_x\phi(x,y)\right)^2+\left(\partial_y\phi(x,y)\right)^2+V^{(2)}(\sl f(x))\phi^2(x,y)}{2}.
\nonumber
\eea

Expanding in terms of $k$ and $p$, integrating by parts and dropping the boundary terms, which are irrelevant for $m>0$, one may simplify the $\phi$ terms
\bea
&&\int_{-\infty}^{\infty} dx \int_{-\pi R}^{\pi R} dy\left[\left(\partial_x\phi(x,y)\right)^2+\left(\partial_y\phi(x,y)\right)^2+V^{(2)}(\sl f(x))\phi^2(x,y)\right]\\
&&\ \ =\frac{1}{4\pi^2R^2}\ppin{k_1}\ppin{k_2}\sum_{p_1,p_2}\phi_{k_1p_1}\phi_{k_2p_2}\int_{-\infty}^{\infty} dx \int_{-\pi R}^{\pi R} dy \g_{k_1}(x)e^{-iy(p_1+p_2)/R}\nonumber\\
&&\ \ \ \times\left(-\partial_x^2-\frac{p_1p_2}{R^2}+V^{(2)}(\sl f(x))\right)\g_{k_2}(x)\nonumber\\
&&\ \ =\frac{1}{2\pi R}\ppin{k_1}\ppin{k_2}\sum_{p_1,p_2}\phi_{k_1p_1}\phi_{k_2p_2}\int_{-\infty}^{\infty} dx \g_{k_1}(x)\delta_{p_1,-p_2}\left(-\frac{p_1p_2}{R^2}+\omega^2_{k_2}\right)\g_{k_2}(x)\nonumber\\
&&\ \ =\frac{1}{2\pi R}\ppin{k}\sum_{p}\Omega^2_{kp}\phi_{kp}\phi_{-k,-p} \hsp
\Omega_{kp}=\sqrt{\omega^2_{k}+\frac{p^2}{R^2}}.
\nonumber
\eea
The $\pi^2$ terms can similarly be simplified
\beq
\int_{-\infty}^{\infty} dx \int_{-\pi R}^{\pi R} dy\ \pi^2(x,y)=\frac{1}{2\pi R}\ppin{k}\sum_{p}\pi_{kp}\pi_{-k,-p}.
\eeq
We then conclude that the free part of the soliton Hamiltonian is a sum of harmonic oscillators
\beq
H\p_2=\frac{1}{4\pi R}\ppin{k}\sum_{p}\left(:\pi_{kp}\pi_{-k,-p}:+\Omega^2_{kp}:\phi_{kp}\phi_{-k,-p}:
\right).
\eeq
This can easily be rewritten in terms of our creation and annihilation operators
\beq
H\p_2=\frac{1}{4\pi R}:\pi^2_{B0}:+\ppin{k}\widehat{\sum}_{p}\Omega_{kp}:B^\ddag_{kp}B_{kp}:
\eeq
where the hat over the sum means that $(k,p)=(B,0)$ is omitted from the sum.

The normal ordering leads to commutator terms, which are $c$-numbers.  They integrate to an infinite quantity.  However, we may reorder instead with all $B^\ddag$ and $\phi_{B0}$ to the left.  This shifts by another $c$-number.  The sum of these two $c$-numbers is $Q_1=2\pi\rho_1$ \cite{noi3d}, which is finite.  We conclude
\beq
H\p_2=Q_1+\frac{1}{4\pi R} \pi^2_{B0}+\ppin{k}\widehat{\sum}_{p}\Omega_{kp}B^\ddag_{kp}B_{kp}. \label{h2}
\eeq
This is the sum of easily solved quantum mechanical systems.  The $\pi_{B0}^2$ term is the Newtonian kinetic energy for the center of mass, while the other terms are quantum harmonic oscillators for the normal modes.  There is therefore a unique ground state
\beq
H\p_2\vac_0=Q_1\vac_0
\eeq
which is defined by
\beq
\pi_{B0}\vac_0=B_{kp}\vac_0=0 \label{vceq}
\eeq
where $k$ runs over the continuum modes $k$ as well as the shape modes $S$ and the zero mode $B$, however $(k,p)=(B,0)$ is omitted as in fact there is no operator $B_{B0}$.
Eq.~(\ref{vceq}) is the statement that there is no center of mass motion and the normal modes are in their ground states.

Recall that the true ground state $\vac$, which is an eigenstate of $H\p$, can be constructed perturbatively, beginning with $\vac_0$ and the domain wall state is $|K\rangle=\df\vac$.  Below we will often omit the $\df$ prefactor to avoid clutter.

\subsection{Interpreting the Zero Modes}

Let us introduce another state, $|x\rangle$, which is similar to $\vac_0$ but is defined by
\beq
B_{kp}|x\rangle=0\hsp \phi_{B0}|x\rangle=x|x\rangle\hsp \langle x|x\rangle=1.
\eeq
Note that, up to an arbitrary factor of proportionality, our nonnormalizable ground state is
\beq
\vac_0=\int_{-\infty}^{\infty} dx|x\rangle.
\eeq
So our ground state is built from these $|x\rangle$ states.  What are they?

Renaming $x=\epsilon$, let us compute the expectation value of the $\phi(x,y)$ field in the state $\df|\epsilon\rangle$
\bea
\frac{\langle \epsilon|\df^\dag \phi(x,y)\df|\epsilon\rangle}{\langle \epsilon|\epsilon\rangle}&=&\frac{\langle \epsilon|\left(\phi(x,y)+f(x)\right)|\epsilon\rangle}{\langle \epsilon|\epsilon\rangle}\\
&=&f(x)+\frac{\langle \epsilon|\left(\frac{1}{2\pi R}\phi_{B0}\g_B(x)\right)|\epsilon\rangle}{\langle \epsilon|\epsilon\rangle}=f(x)+\frac{\epsilon}{2\pi R}\left(-\frac{f\p(x)}{\sqrt{\rho_0}}\right)\nonumber\\
&=&f\left(x-\frac{\epsilon}{2\pi R\sqrt{\rho_0}}\right)+O(\epsilon^2).\nonumber
\eea
In other words, the state $\df|\epsilon\rangle$ represents a domain wall string at a position $\epsilon/(2\pi R\sqrt{\rho_0})$.  It is not a Hamiltonian eigenstate, it is a position eigenstate.  To obtain a Hamiltonian eigenstate, one must sum over all positions.  Therefore the ground state domain wall string is delocalized in the $x$ direction.  This is consistent with the usual intuition that the translation symmetry cannot be spontaneously broken if the orthogonal direction is compact.

We have learned that the eigenvalue of $\phi_{B0}$, divided by the dimensionless quantity $2\pi R\sqrt{\rho_0}$, is the position of the center of mass of the domain wall.  From the commutation relation
\beq
[\phi_{B0},\pi_{B0}]={2\pi i R} \label{zc}
\eeq
we learn that the eigenvalue of $\pi_{B0}$ must be its momentum divided by $\sqrt{\rho_0}$.  Note that the leading order mass is $Q_0=2\pi R\rho_0$ and so the Newtonian kinetic energy is 
\beq
\frac{P^2}{2Q_0}=\frac{\left(\sqrt{\rho_0}\pi_{B0}\right)^2}{4\pi R\rho_0}=\frac{\pi^2_{B0} }{4\pi R}
\eeq
which is indeed the kinetic term reported in Eq.~(\ref{h2}).

\section{The Domain Wall String on a Big Cylinder} \label{gsez}

\subsection{Zero Modes}

Intuitively, we will be interested in an infinitely wide cylinder.  We would like to take the limit that $R$ is infinite, in order to eventually understand the uncompactified case.  Of course $R$ is dimensionful, and so this limit must be defined carefully.  To do this, we will introduce a timescale $T$.  We will only be interested in the initial value problem involving time evolution of times less than $T$ and experiments whose results can be derived from this problem.  Large $R$ will mean $T\ll R$.  However, we also wish $T$ to be large.  Therefore, we will take the double scaling limit in which $T$ is larger than our other scales.  For example, we will take the limit $\lambda^jm^{1-j} T\rightarrow\infty$ for all fixed $j$.  In other words, $T$ is large enough for any perturbative process to occur.  In summary, we are interested in the double scaling limit
\beq
\frac{T}{R}\rightarrow 0\hsp \lambda^jm^{1-j} T\rightarrow\infty.
\eeq

Energy differences are observable because, over a time $T$, they lead to a phase shift $ET$.  As a result, any two states whose energies differ by $E$ such that $ET\rightarrow 0$ in our limit will be degenerate.

Let us first consider the zero mode contribution to the energy.  Multiplied by $T$, this is
\beq
\frac{ T }{4\pi R}\pi^2_{B0}.
\eeq
There will be no contribution to the energy of a state if the expectation value of this quantity vanishes in our limit.  

When does this happen?  Let us consider a real-valued function $\psi$ and a state
\beq
|\psi\rangle=\int_{-\infty}^{\infty} dx \psi(x)|x\rangle.
\eeq
Using the commutator (\ref{zc}) one finds
\beq
\pi_{B0}|\psi\rangle=-2\pi i{R}\int_{-\infty}^{\infty} dx \psi\p(x)|x\rangle.
\eeq 
Then the expectation value is 
\beq
\frac{T }{4\pi R} \frac{\langle\psi|\pi^2_{B0}|\psi\rangle}{\langle \psi|\psi\rangle}=\pi{RT}\frac{\langle\psi\p|\psi\p\rangle}{\langle \psi|\psi\rangle}\ll \pi{R^2}\frac{\langle\psi\p|\psi\p\rangle}{\langle \psi|\psi\rangle}=\pi R^2\frac{\int_{-\infty}^{\infty} dx \psi^{\prime 2}(x)}{\int_{-\infty}^{\infty} dx \psi^2(x)}.
\eeq
If the derivative scale $\psi\p(x)/\psi(x)$ is of order $O(1/R)$ or less, this vanishes in our large radius limit.  Now the position of a string is not $x$, it is {$x/(2 \pi R\sqrt{\rho_0})$} and so a more useful form for the condition is
\beq
\frac{1}{\psi(x)}\frac{\partial}{\partial (x/(R\sqrt{\rho_0})} \psi(x)\lesssim\sqrt{\rho_0)}.
\eeq

At finite $R$ we saw that $\psi(x)$ needs to be a constant, corresponding to a completely homogeneous zero mode.  Now, we see that, in our double scaling limit, there are many ground states corresponding to different functions $\psi(x)$.  The function $\psi(x)$ no longer needs to be flat, but it can have a derivative so long as the derivative scale does not exceed $\sqrt{\rho_0}$ by any powers of $R$.  It may, however, exceed $\sqrt{\rho_0}$  by powers of inverse coupling, and so intuitively any function $\psi(x)$ which is differentiable in the large $R$ limit will provide a degenerate vacuum.  
In particular, the zero mode wave function apparently can break translation symmetry with no cost in energy.  

\subsection{Nonzero Modes and Coleman's Theorem}

On the worldsheet of the 1+1 dimensional domain wall string, the embedding coordinate $x_0$ is a field and the translation is an internal symmetry.  Of course we know \cite{colemanth} that continuous internal symmetries cannot be broken in 1+1 dimensions, and so it would be rather surprising if the translation symmetry could be broken here. 
So let us press on and study the $x$ distribution caused by the nonzero modes.

For brevity, as the focus of our paper is the zero modes, in this subsection we will ignore $T$ and simply place the nonzero modes in their ground states.  We relegate the question of whether there are more general choices of ground state for the nonzero modes to future work.  We consider this to be quite unlikely, as given any fixed nonzero mode, there is a sufficiently large $T$ such that it is arbitrarily close to its ground state.

We have seen in Eq.~(\ref{vceq}) that the ground state $\vac_0$ is annihilated by each annihilation operator $B_{kp}$, where we recall that no such operator exists for $(k,p)=(B,0)$.  Let us write the operator as
\beq
B_{-k,-p}=\sqrt\frac{1}{2\pi R}\left(\Omega_{kp}\phi_{kp}+i\pi_{kp}\right)=\sqrt\frac{1}{2\pi R}\Omega_{kp}\phi_{kp}+\sqrt{2\pi R}{}\frac{\partial}{\partial\phi_{kp}}.
\eeq

We can write the domain wall string ground state in terms of the wave functions of $\phi_{kp}$.  In other words, let us define the $\phi_{kp}$ eigenstates
\beq
\phi_{kp}|x\rangle_{kp}=x|x\rangle_{kp}.
\eeq
Then, for any real-valued function $\psi_{kp}(x)$ we can define the wave packet
\beq
|\psi_{kp}\rangle=\int_{-\infty}^{\infty} dx \psi_{kp}(x)|x\rangle_{kp}.
\eeq
This is in the kernel of the annihilation operator if
\beq
0=B_{-k,-p}|\psi_{kp}\rangle=\sqrt\frac{1}{2\pi R}\int_{-\infty}^{\infty} dx\left[\Omega_{kp}x\psi_{kp}(x)+2\pi{R}{}\psi\p_{kp}(x) 
\right]|x\rangle_{kp}.
\eeq
Since the $|x\rangle_{kp}$ are linearly independent, the term in the square brackets must vanish
\beq
\psi\p_{kp}(x)=-\frac{1}{2\pi R}\Omega_{kp}x\psi_{kp}(x).
\eeq
We then learn that, up to a constant, the wave function is
\beq
\psi_{kp}(x)=\exp{-\frac{1}{4\pi R}\Omega_{kp}x^2}.
\eeq

Physically this means that the eigenvalue $x$, of $\phi_{kp}$ has a Gaussian distribution with a variance of $2\pi R/\Omega_{kp}$.  Consider, for example, the case of the translation modes, $k=B$.  Then the distribution of eigenvalues of $\phi_{Bp}$ has a variance of $2\pi R^2/|p|$.

We have seen that the position of the string is the eigenvalue of $\phi_{B0}$ divided by $2\pi R\sqrt{\rho_0}$.  The eigenvalue of $\phi_{Bp}$ has the same interpretation, but with an additional factor of $e^{-ipy/R}$.  For simplicity, let us consider $y=0$ so that this factor is unity.  $y$-translation invariance implies that this choice can be made without loss of generality.  At the position $y=0$, we conclude that as a result of the $\phi_{Bp}$ mode, the variance of the $x$ position is
\beq
\Delta x^2_{Bp}=\frac{2\pi R^2}{|p|}\left(\frac{1}{2\pi R\sqrt{\rho_0}}\right)^2=\frac{1}{2\pi \rho_0}\frac{1}{|p|}.
\eeq

The total wave function for $\vac_0$ is the tensor product of all of these factors, together with that of the zero mode.  Convoluting Gaussians, one arrives at another Gaussian whose variance is the sum of that of the original Gaussians. Therefore, the total variance in the position resulting from the modes $k=B$ with $p\neq 0$ is
\beq
\Delta x^2= \frac{1}{2\pi \rho_0}\sum_{p\neq 0}\frac{1}{|p|}. \label{disc}
\eeq
This sum diverges, and so one concludes that the $x$ distribution is a Gaussian of infinite variance.  In other words, it is a flat distribution, and so the translation-invariance is not broken, in accordance with Coleman's theorem.

One needs to be careful here.  Coleman's theorem states that the translation-invariance is unbroken in the infrared.  But this sum appears to be divergent at large $p$, corresponding to the ultraviolet.  

Is there a divergence in the infrared?  Recall that at finite $R$ the zero mode $\phi_{B0}$ already restored the translation-invariance that was broken by the classical solution.  Thus, the question is only relevant in our large $R$ limit.  We will use $\bp$ to indicate the momentum and an ordinary $p$ to be the discrete index which labeled the momentum on a compact space.  These are related by
\beq
\bp=\frac{p}{R}\hsp \pin{\bp}=\frac{1}{2\pi R}\sum_{p}.
\eeq
So we find
\beq
\Delta x^2= \frac{1}{2\pi \rho_0}\sum_{p\neq 0}\frac{1}{|p|}=\frac{1}{2\pi\rho_0}\int d\bp\frac{1}{|\bp|}. \label{div}
\eeq
This integral indeed contains, in addition to the ultraviolet divergence, an infrared divergence at small ${\bp}$.  The infrared divergence was already present in Eq.~(\ref{disc}) as the observation that the momentum is $p/R$ and so, at large $R$, infinitely many values of $p$ are in the deep infrared.  This infinity of values is divergent when summed.

\section{A Noncompact Domain Wall String} \label{infsez}

Finally, we turn to the domain wall string in infinite Minkowski space.  Of course, we took the limit in which $R$ was infinite in the previous section.  So in this section we will define the domain wall string in Minkowski space to be that of the previous section, in other words that obtained from the domain wall string on a circle in the limit described above.  Again using ${\bp}=p/R$, we may write the components of the fields with continuous momenta ${\bp}$ in terms of those with discrete momenta $p$ as
\beq
\phi_{k\bp}=\phi_{k,p/R}\hsp \pi_{k\bp}=\pi_{k,p/R}.
\eeq
These satisfy the canonical commutation relations
\beq
[\phi_{k_1{\bp}_1},\pi_{k_2{\bp}_2}]=\left\{\begin{tabular}{lll}
$i\ 2\pi\delta({\bp}_1+{\bp}_2)\delta_{k_1k_2}$&&\rm{if $k_i$ are discrete}\\
$i\ 2\pi\delta({\bp}_1+{\bp}_2) 2\pi\delta(k_1+k_2) $&&\rm{if $k_i$ are continuous}.
\end{tabular}
\right.
\eeq

In terms of these, the continuum limit of the decomposition of the fields is
\beq
\phi(x,y)=\ppin{k}\pin{\bp} \phi_{k\bp}\g_k(x)e^{-iy\bp}\hsp
\pi(x,y)=\ppin{k}\pin{\bp} \pi_{k\bp}\g_k(x)e^{-iy\bp}.\label{cdec}
\eeq

Note that the commutator of the zero modes $\phi_{B0}$ and $\pi_{B0}$, which was $2\pi i R$ in the discrete case, is infinite in the continuum limit.  Moreover, the position of the string is the eigenvalue of $\phi_{B0}$ divided by the infinite quantity $2\pi R\sqrt{\rho_0}$.  Both problems are resolved if we define the operator \beq
\phi_0=\frac{\phi_{B0}}{2\pi R}.
\eeq
How can this be defined in our limit?

Using the fact that
\beq
\phi_{k0}=\int dx \g_{-k}(x)\int dy \phi(x,y)
\eeq
we see that instead of just integrating over $y$, we need to average over $y$
\beq
\phi_0=\lim{R\rightarrow\infty} \frac{1}{2\pi R} \int dx \g_{B}(x)\int_{-\pi R}^{\pi R} dy \phi(x,y).
\eeq
Then we obtain the finite commutator
\beq
[\phi_0,\pi_{B0}]=i. \label{fcom}
\eeq
Also, the expectation value of $\phi_0$ can be identified with the position of the string divided by the finite quantity $\sqrt{\rho_0}$.

In terms of $\phi_0$, our decomposition can be written
\beq
\phi(x,y)=\phi_0\g_B(x)+\frac{1}{2\pi R}\ppin{k}\widehat{\sum_{p}}\phi_{kp}\g_k(x)e^{-iyp/R}.
\eeq
In the continuum limit, 
\beq
\phi(x,y)=\phi_0\g_B(x)+\ppin{k}\dashint\frac{d\bp}{2\pi} \phi_{k\bp}\g_k(x)e^{-iy\bp}
\eeq
where, in the second term, when $k=B$, the $\bp$ integration omits the origin.  The first term, on the other hand, is that of the kink in Ref.~\cite{mekink}.  In particular, it implies that the interactions $H\p_{i\geq 3}$ will include factors of the zero mode $\phi_0$, just as in the case of the kink.  

However, unlike the case of the kink, these factors are not easily canceled by the center of mass kinetic term $\pi_{B0}^2/(4\pi R)$ in $H\p_2$, which was shown in Eq.~(\ref{h2}).  Consider a state $|\psi\rangle$ with no center of mass momentum, so that
\beq
\pi_{B0}|\psi\rangle=0.
\eeq
Then commutator of the center of mass term with $\phi^2_0$, acting on  $|\psi\rangle$, is 
\beq
\left[ \frac{\pi_{B0}^2}{4\pi R},\phi_0^2\right]|\psi\rangle=-\frac{|\psi\rangle}{2\pi R}.
\eeq
This is in contrast with the case of the kink, in which the commutator of the kinetic term $\pi_0^2/2$ and $\phi_0^2$, when acting on $|\psi\rangle$, is equal to $-1$. In the case of the kink, this means that interactions create zero modes which are eliminated by the free evolution in $H\p_2$, leading to contributions of order unity to mass shifts, scattering amplitudes and other observables.  In the present case, interactions still create zero modes $\phi_0$ but these zero modes cannot be removed by the free evolution without a factor of $1/(2\pi R)$.  Thus, these zero mode contributions, which played a prominent role in the case of the kink, do not appear in the case of extended solitons.  Therefore we expect that these zero modes will not contribute to the scattering of domain walls with finite numbers of perturbative mesons, unlike the case of kinks in Ref.~\cite{elastic}.  However, we leave the verification of this conjecture to a study of loop corrections to such scattering amplitudes in future work.

There is one way to escape this argument, the initial interaction may be itself amplified by a power of the radius.  This will be the case, for example, if a constant pressure is exerted on a domain wall, for example if the vacuum energies on the two sides are different. 

Note that although there are now degenerate ground states, they all have the same energy.  Indeed that is how we defined the ground state.  However, in principle there is a zero mode contribution to $H\p_2$.  To take the continuum limit of $H\p_2$, first we will define the continuum creation and annihilation operators
\beq
B_{k\bp}=\sqrt{2\pi R}B_{k,p/R}\hsp B^\ddag_{k\bp}=\sqrt{2\pi R}B^\ddag_{k,p/R}.
\eeq
which satisfy the algebra
\beq
[B_{k_1 \bp_1},B^\ddag_{k_2\bp_2}]=\left\{\begin{tabular}{lll}
$2\pi\delta({\bp}_1-{\bp}_2)\delta_{k_1k_2}$&&\rm{if $k_i$ are discrete}\\
$2\pi\delta({\bp}_1-{\bp}_2)2\pi\delta(k_1-k_2)$&&\rm{if $k_i$ are continuous.}
\end{tabular}
\right.
\eeq
One then finds
\beq
H\p_2=Q_1+\frac{\pi_{B0}\pi_0}{2}
+\ppin{k}\dashint\frac{d\bp}{2\pi} \Omega_{k\bp}B^\ddag_{k\bp}B_{k\bp}
\eeq
where we have defined
\beq
\pi_0=\lim{R\rightarrow\infty} \frac{1}{2\pi R} \int dx \g_{B}(x)\int _{-\pi R}^{\pi R} dy \pi(x,y).
\eeq
Note that the commutator of $\pi_0$ and $\phi_0$ vanishes in the continuum limit.  However, this zero is parameterized, it results from division by the length of the domain wall string.  Therefore, loop effects involving the creation of two zero modes $\phi_0$ followed by their annihilation by this kinetic term will be suppressed by the length of the string.  As a result, they may contribute to amplitudes that are enhanced by the length of the string, such as the effect of an external source, radiation field or vacuum condensate which acts coherently over the entire, infinite string length.  Indeed, one expects that an infinite string can be moved if subjected to a constant pressure.

What are the ground states of the continuum string?  Again, the Schrodinger wave functional factorizes into a tensor product $\vac_+$ of the ground states of the various oscillator terms, which are Gaussians so as to be annihilated by each $B_{k\bp}$, and a wave function $\psi(\phi_0)$ for the zero mode.  In the case of the domain wall string, unlike higher-dimensional domain walls, this zero mode wave function cannot localize the string because the nonzero modes delocalize it.  Nonetheless, it is a part of the wave function and describes a set of degenerate ground states.  Schematically, we may write each ground state $\vac_\psi$ as
\beq
\vac_\psi=\vac_+\otimes \int dx \psi(x) |x\rangle\hsp B_{k\bp}\vac_+=0\hsp \phi_0|x\rangle=x|x\rangle.
\eeq
Note that the different states $|x\rangle$ are related by an infinite action instanton, as a result of the infinite length of the $y$ direction.  Therefore they represent distinct superselection sectors.  As a result, in the absence of infinite sources, one expects that perturbative amplitudes will not be sensitive to any interference between these sectors.  Therefore, if there are no impurities breaking the $x$-translation invariance, the superselection sectors will be identical and so the choice of wavefunction will not affect the computation of any observables.  In particular, we expect that the choice of wave function will not affect the amplitudes for any perturbative process, and so one can generally safely ignore this component of the wave function.  Nonetheless, as $\phi_0$ appears unsuppressed in the decomposition of $\phi(x,y)$ and so in interactions $H\p_{i\geq 3}$, this naive intuition needs to be tested.

\section{The Domain Wall Soliton} \label{memsez}

This construction is easily generalized to the domain wall membrane in 3+1 dimensions.  One merely replaces each $2\pi R$ with a volume factor.  There is little difference, except for the sum of the variances in Eq.~(\ref{disc}) which must now be taken over a two-dimensional lattice.   This introduces a factor of ${\bp}$ in the numerator of Eq.~(\ref{div}) which removes the infrared divergence.  As a result, the variance is finite and there are ground states in which the domain wall is localized.  Indeed, Coleman's theorem does not apply to (2+1)-dimensional worldvolumes.

\section{Stokes Scattering} \label{stsez}

\subsection{Calculating Amplitudes in General}

Unlike the case of the quantum kink, there is not a unique ground state string.  Rather, the ground state has a degeneracy given by a factor $\int dx \psi(x) |x\rangle$.  This appears not only in the ground state, but also in excited states and in states containing perturbative mesons in addition to the string.

If the wave function $\psi$ does not evolve during a physical process, then clearly it will not affect the amplitude.  However, more generally, one may consider $x$, the center of mass position of the string, to be a new quantum number.  Then, in assembling the probabilities into an amplitude, distinct values of $x$ need to be summed incoherently.  In other words, one should calculate the probability for each $x$ by squaring the amplitude and then sum over the probabilities.  Therefore if the wave function $\psi(x)$ changes, the probabilities may also change.  In particular, if the interaction Hamiltonian leads to an evolution operator that multiplies a state by $\phi_0$, the interference of this channel with a channel that does not contain the factor of $\phi_0$ will affect the probability.  Therefore it is important to consider such terms in the evolution.

\subsection{Stokes Scattering}

At leading order, an eigenstate of the wall Hamiltonian $H\p$ is an eigenstate of $H\p_2$.  We will consider Stokes scattering in 2+1 dimensions, in which an incoming meson strikes a domain wall string in its ground state, exciting its translation $(B\bp_3)$ mode and either continuing through the string or else reflecting back.  In other words, incoming radiation creates a displacement wave on the string with a momentum of $\bp_3$.  This is in contrast with Ref.~\cite{hengyuan}, which treated the process in which a shape mode was created.

The $H_2\p$ eigenstate corresponding to a ground state string and an incoming meson is
\beq
|k_1,\bp_1\rangle_{0,\psi}=B^\ddag_{k_1\bp_1}\vac_\psi
\eeq
where $\psi$ is a wave function for the zero mode.  Here $B^\ddag_{k_1\bp_1}$ is the creation operator for a free meson whose momentum in the $x$ direction is $k_1>0$ and whose $y$-momentum is $\bp_1$.  The final state of interest consists of a meson and a string with an excited translation mode with $y$-momentum $\bp_3$.  The corresponding $H\p_2$ eigenstate is
\beq
|k_2,\bp_2;B\bp_3\rangle_{0,\psi_2}=B^\ddag_{k_2\bp_2}B^\ddag_{B\bp_3}\vac_{\psi_2}
\eeq
where $k_2$, $\bp_2$ and $\bp_3$ are real numbers corresponding to the $x$ and $y$ momenta of the outgoing meson and the $y$ momentum of the outgoing translation mode respectively.  Here $\psi_2$ is another arbitrary wave function for the zero mode.

At leading order, these are connected by the  interaction
\beq
H\p_3=\frac{\sl}{6}\int dx {V^{(3)}(\sqrt{\lambda} f(x))}\int dy: \phi^3(x,y):.
\eeq
More precisely, they are connected by the term
\bea
H_I&=&{\sqrt{\lambda}}{}\int\frac{d^2k d^3\bp}{(2\pi)^5}\frac{V_{-k_1,-\bp_1;k_2\bp_2;B\bp_3}}{2\Omega_{k_1\bp_1}}B_{B\bp_3}^{\ddagger}B^\ddag_{k_2\bp_2}B_{k_1 \bp_1}\label{hint}\\
V_{-k_1,-\bp_1;k_2\bp_2;B\bp_3}&=&2\pi\delta(\bp_1-\bp_2-\bp_3)V_{B,k_2,-k_1}\nonumber\\
V_{B,k_2,-k_1}&=&\int dx V^{(3)}(\sqrt{\lambda} f(x))\g_B(x)\g_{k_2}(x)\g_{-k_1}(x). \nonumber
\eea 
Note that this interaction does not contain the zero mode $\phi_0$, and so the zero mode part of the wave packet will not evolve during the leading order process and we may restrict our attention to the case $\psi_2=\psi$.  With this simplification at tree level, we conclude that our calculation will mirror that of Ref.~\cite{hengyuan} quite closely.

We will begin our Stokes scattering with the initial condition
\bea
|\Phi\rangle&=&\pin{\bp_1}\a_{k_1\bp_1}|k_1,\bp_1\rangle_{0,\psi}\hsp
\alpha_{k_1\bp_1}=\int dx\int dy \Phi(x,y) \g_{k_1}(x)e^{-i\bp_1 y}\\
\Phi(x,y)&=&e^{-\frac{(x-x_0)^2}{4\sigma^2}+ik_0x}e^{-\frac{y^2}{4\sigma^2}}\nonumber
\eea
where the initial position $x_0$, $x$-momentum $k_0$ and wave packet width $\sigma$ satisfy
\beq
x_0 \ll -\frac{1}{m} \hsp 0<\frac{1}{k_{0}},\frac{1}{m} \ll \sigma \ll |x_0|.
\eeq
Here $\Phi(x,y)$ is the position space incoming meson wave packet.  We have not considered a center of mass $y$-momentum $\bp_0$ for the initial wave packet, but that case may be obtained from the present case via a Lorentz transformation.

Far from the string, the normal modes are
\bea
\g_{k}(x)&=&\left\{\begin{tabular}{lll}
$\mb_{k}e^{-ik x}+\mc_{k}e^{ik x}$  &\rm{if} & $x\ll  -1/m$\\
$\md_{k}e^{-ik x}+\me_{k}e^{ik x} $  &\rm{if} & $x\gg 1/m$\\
\end{tabular}
\right. \label{gk}\\
|\mb_{k}|^2+|\mc_{k}|^2&=&|\md_{k}|^2+|\me_{k}|^2=1\hsp
\mb^*_{k}=\mb_{-k}\hsp
\mc^*_{k}=\mc_{-k}\hsp
\md^*_{k}=\md_{-k}\hsp
\me^*_{k}=\me_{-k}\nonumber
\eea
where $\mc$ and $\me$ can be taken to vanish for a reflectionless potential such as in the $\phi^4$ model.  

Let us introduce the function $k_I(k_2,\bp_2,\bp_3)$ which is defined by
\beq
k_I(k_2,\bp_2,\bp_3)>0\hsp \Omega_{k_I(k_2,\bp_2,\bp_3),\bp_2+\bp_3}=\Omega_{k_2\bp_2}+|\bp_3|.
\eeq
In other words, for a given set of outgoing momenta, $k_I$ is the incoming $x$-momentum that would make the process on shell.  Below we will omit the arguments of $k_I$ to avoid clutter.

At order $O(\sl)$, the Schrodinger picture evolution via $H_I$ proceeds as in Ref.~\cite{hengyuan} leading at large times $t$ to a state whose 1-meson, 1-translation mode component is
\bea
e^{-iH\p t}|\Phi\rangle\bigg|_{O(\sl)}
&=&-i\sigma\sqrt{\pi\lambda}\int\frac{dk_2d\bp_2d\bp_3}{(2\pi)^3}e^{ix_0(k_{0x}-k_I)}e^{-\sigma^2(k_{0}-k_I)^2}e^{-i{(\Omega_{k_2\bp_2}+|\bp_3|)}t}\nonumber\\
&&\times e^{-\sigma^2({\bp_2+\bp_3})^2}\left(\frac{\tilde{V}_{Bk_{2},-k_I}}{k_I}\right)|k_2\bp_2,B\bp_3\rangle_{0,\psi}\nonumber\\
\tilde{V}_{Bk_{2},-k_I}&=&\mb_{k_I}V_{B,k_{2},-k_I}+\mc^*_{k_I}V_{B,k_{2},k_I}.  
\eea
Note that in the case of a reflectionless wall, such as in the $\phi^4$ model, $\mc=0$ and so $\tilde{V}=V$.  One can see that the zero mode wave function, $\psi$, is unchanged as the only contribution to this component of the state at $O(\sl)$ did not include any zero mode operators $\phi_0$ or $\pi_0$.

Following the calculation in Ref.~\cite{hengyuan} this leads to the differential probability of Stokes scattering given an incoming momentum $k_0$
\beq
\frac{dP_{\rm{S}}(k_0,\bp_3)}{d \bp_3}
=\frac{\lambda\left(\left|\tilde{V}_{B,\sqrt{k_0^2-2\omega_{k_0}|\bp_3|},-k_0}\right|^2
+\left|\tilde{V}_{B,-\sqrt{k_0^2-2\omega_{k_0}|\bp_3|},-k_0}\right|^2\right)}{16\pi|\bp_3|k_0\sqrt{k_0^2-2\omega_{k_0}|\bp_3|}}
.\label{psm}
\eeq
Note that the differential probability naively diverges as $1/|\bp_3|$ when the $y$-momentum of the zero mode is small.  Such an infrared divergence for soft zero modes itself is not necessarily problematic, however the semiclassical expansion itself breaks down when this probability is of order unity.  This implies that this contribution to the probability does not dominate if $|\bp_3|\lesssim m\lambda$, and conversely the range of validity of our results is naively $|\bp_3|\gg m\lambda$.  

\subsection{The $\phi^4$ Double-Well}

Let us restrict our attention to the $\phi^4$ double-well model, defined by the potential.
\beq
V(\sqrt{\lambda}\phi(x))=\frac{\lambda\phi^2(x)}{4}\left(\sqrt{\lambda}\phi(x)-\sqrt{2}m\right)^2.
\eeq
Classically, it exhibits a domain wall solution
\beq
f(x)=\frac{m}{\sqrt{2\lambda}}\left(1+\tanh\left(\frac{mx}{2}\right)\right). \label{f}
\eeq
This domain wall enjoys continuous normal modes, a shape mode and a zero-mode which are respectively
\bea
\g_k(x)&=&\frac{e^{-ikx}}{\ok{} \sqrt{m^2+4k^2}}\left[2k^2-m^2+(3/2)m^2\sech^2(m x/2)-3im k\tanh(m x/2)\right]\nonumber\\
\g_S(x)&=&\frac{\sqrt{3 m}}{2}\tanh(m x/2)\sech(m x/2)\hsp
\g_B(x)=-\sqrt{\frac{{3m}}{8}}\sech^2(m x/2).\label{nmode}
\eea
Combining these, one finds the quantity $V_{Bkk}$ defined in Eq.~(\ref{hint})
\beq
V_{Bk_1k_2}=i\pi \frac{3\sqrt{3}}{m^{3/2}}\frac{\left(\ok1^2-\ok2^2\right)^2(m^2+k_1^2+k_2^2)}{\ok1\ok2\sqrt{m^2+4k_1^2}\sqrt{m^2+4k_2^2}}\csch\left(\frac{\pi(k_1+k_2)}{m}\right). \label{vbkk}
\eeq
The $\phi^4$ domain wall is reflectionless, and so $\tilde{V}=V$.  Substituting (\ref{vbkk}) into our master formula Eq.~(\ref{psm}) for the Stokes scattering differential probability, we find
\bea
\frac{dP_{\rm{S}}(k_0,\bp_3)}{d \bp_3}
&=&\frac{\lambda\left(\left|\tilde{V}_{B,\sqrt{k_0^2-2\omega_{k_0}|\bp_3|},-k_0}\right|^2
+\left|\tilde{V}_{B,-\sqrt{k_0^2-2\omega_{k_0}|\bp_3|},-k_0}\right|^2\right)}{16\pi|\bp_3|k_0\sqrt{k_0^2-2\omega_{k_0}|\bp_3|}}\\
&=&\frac{27\lambda \pi \omega_{k_0}^2 |\bp_3|^3  (m^2+2k_0^2-2\omega_{k_0}|\bp_3|)^2}{m^3 k_0 (m^2+4k_0^2)\left(m^2+4k_0^2-8\omega_{k_0}|\bp_3|\right) (m^2+k_0^2-2\omega_{k_0}|\bp_3|)\sqrt{k_0^2-2\omega_{k_0}|\bp_3|}}
\nonumber\\
&&\times \left[\csch^2\left(\frac{\pi(\sqrt{k_0^2-2\omega_{k_0}|\bp_3|}-k_0)}{m}\right) +\csch^2\left(\frac{\pi(\sqrt{k_0^2-2\omega_{k_0}|\bp_3|}+k_0)}{m}\right)\right]. \nonumber
\eea
Note that there is no infrared divergence at small $\bp_3$.  The naive divergence has been canceled by the $\bp_3^2$ term in $(\ok{1}^2-\ok{2}^2)$ which appeared in each $V_{BBk}$.  There is still the usual kinematic divergence at small $k_0$, arising from the fact that the meson spends more time close to the wall and so is more likely to interact.

\begin{figure}[htbp]
\centering
\includegraphics[width = 0.65\textwidth]{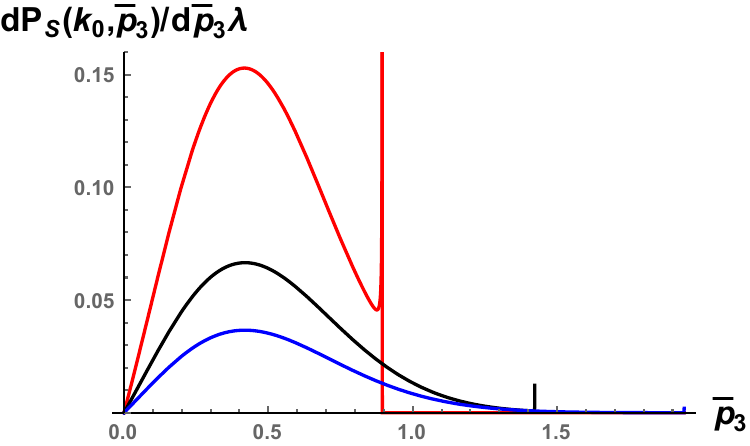}
\caption{The probability density for the creation of a translation mode with  momentum $\overline{p}_3$ when the incoming meson has momentum $k_0$ equal to $2m$ (red), $3m$ (black) and $4m$ (blue).  The scale is fixed by the convention $m=1$.}\label{highk}
\end{figure}

\begin{figure}[htbp]
\centering
\includegraphics[width = 0.65\textwidth]{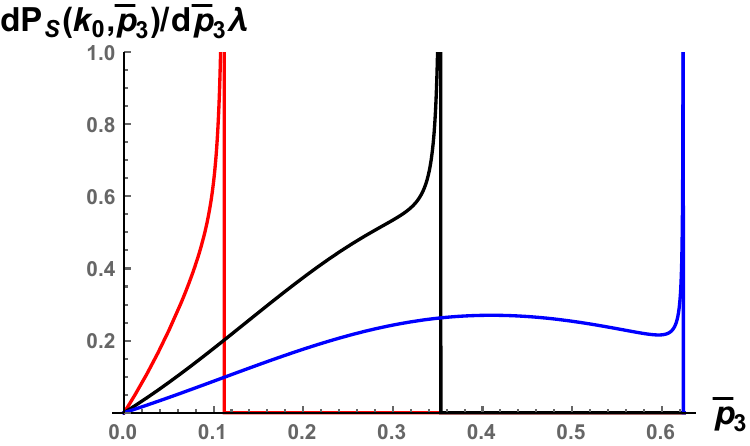}
\caption{The probability density for the creation of a translation mode with  momentum $\overline{p}_3$ when the incoming meson has momentum $k_0$ equal to  $m/2$ (red), $m$ (black) and $3m/2$ (blue).  The scale is fixed by the convention $m=1$.}\label{lowk}
\end{figure}

We plot this probability density for several ultrarelativistic values of the incoming momentum $k_0$ in Fig.~\ref{highk}.  Notice that at $k_0=2m$ there is already a noticeable peak at the maximal momentum transfer $\overline{p}_3$ to the translation mode, with the outgoing meson traveling in the opposite direction parallel to the string.  At lower incoming momentum, shown in Fig.~\ref{lowk}, this threshold peak eventually dominates. 

\section{Remarks}
We have answered most of the questions posed in the introduction.  The zero modes of an extended soliton of finite length, in its ground state, are in a flat superposition as in the case of a localized soliton.  As in the case of the localized soliton, this flat superposition is not normalizable.

On the other hand, we have seen that for an infinitely extended soliton, perhaps because the distinct zero mode eigenvalues correspond to distinct superselection sectors, even in the ground state we have the freedom to choose the zero mode wave function.  We have seen that this choice is compatible with Coleman's theorem as a result of the nonzero modes.  The choice allows the zero mode wave function to be normalizable, simplifying calculations with respect to the case of the localized or compact soliton.

Note that while this observation applies to the ground state, if any finite energy is added then an infinite dimensional space of long-wavelength modes $B^\ddag_{Bp}\vac_0$ is accessible, which potentially can lead to infrared divergences.  Interestingly, in the case of Stokes scattering, we saw that the differential amplitude for exciting such states was suppressed, even when integrating over all such soft modes.  

We have explicitly constructed the ground states of domain wall solitons, including the zero mode part which had been glossed over in previous constructions \cite{noi3d}.  As a result, we were able to calculate the differential probability for the scattering of a meson off a domain wall string which imparts some translation mode to the domain wall string, making it move.  As expected based on our results, the contribution of the zero mode turned out to be rather trivial, and our probability was that which one may have naively guessed by merely replacing the shape mode with the zero mode in the results of Ref.~\cite{hengyuan}.

We have not yet determined whether the zero modes still appear in loops, as they do in the case of localized solitons.  We have constructed the relevant interaction terms.  In the future, we hope to apply these to loop interactions, such as elastic meson-string scattering, to see whether the zero modes contribute.

The present paper treated zero modes arising from spatial translations.  Time-dependent classical solutions, such as Sine-Gordon breathers, also correspond to quantum states.  As in the case of the harmonic oscillator, the classical time dependence corresponds to quantum coherent states, but there are also Hamiltonian eigenstates in the quantum theory.  There are also zero modes corresponding to time translations, and the Hamiltonian eigenstates contain wave functions over these zero modes.  In the case of the oscillon, which in some sense includes the breather as a limiting case, these temporal zero modes have recently been constructed in Ref.~\cite{oscspec}.  It would be interesting to extend the present analysis to these zero modes, to understand how they evolve and to understand various choices of wave functions.  

The relationships between $Q$-balls and oscillons are intricate and rich \cite{qosc1,qosc2,qosc3}.  It would be interesting to understand how the zero modes assemble in these correspondences.

\section*{Acknowledgement}

\noindent
This work was supported by the Higher Education and Science Committee of the Republic of Armenia (Research Project Nos. 24RL-1C047 and 24PostDoc/2‐1C009). This work was also partly supported by a short term scientific mission grant from the COST action CA22113 THEORY-CHALLENGES.


\end{document}